\documentclass{aastex}
\usepackage{spr-astr-addons}
\usepackage{url}

\RequirePackage{color}

\newcommand{\emaila}{authors@email.com}

\begin{document}
\pagestyle{headings}
\title{Radiative properties of stellar plasmas\\and open challenges}
\noindent
 {\bf Published in Astrophysics and Space Science}
\shorttitle{Radiative  properties of stellar plasmas}
\shortauthors{Turck-Chièze, Gilles, Loisel et al.}

\author{S. Turck-Chi\`eze\altaffilmark{1}, G. Loisel \altaffilmark{2,1}, D. Gilles\altaffilmark{1},  L. Piau\altaffilmark{1},
C. Blancard\altaffilmark{3}, T. Blenski\altaffilmark{2},  M. Busquet\altaffilmark{4}, T. Caillaud \altaffilmark{3}, P. Coss\'e\altaffilmark{3},  
F. Delahaye\altaffilmark{5},  G. Faussurier\altaffilmark{3}, J. Fariaut \altaffilmark{3}, F. Gilleron\altaffilmark{3},
J.A. Guzik\altaffilmark{6}, J. Harris\altaffilmark{7}, 
D.P. Kilcrease\altaffilmark{6}, N.H. Magee\altaffilmark{6}, J.C. Pain\altaffilmark{3}, 
Q. Porcherot\altaffilmark{3}, M. Poirier\altaffilmark{2}, G. Soullier\altaffilmark{3}, C.J. Zeippen\altaffilmark{5}, 
S. Bastiani-Ceccotti\altaffilmark{8}, C. Reverdin\altaffilmark{3}, V. Silvert\altaffilmark{3}, F. Thais\altaffilmark{2}, B. Villette \altaffilmark{3}}
\email{\emaila{sylvaine.turck-chieze@cea.fr}}
\altaffiltext{1}{CEA/DSM/IRFU/SAp, CE Saclay, F-91190 Gif sur Yvette, France}
\altaffiltext{2}{CEA/DSM/IRAMIS/SPAM, CE Saclay, F-91190 Gif sur Yvette}
\altaffiltext{3}{CEA/DAM/DIF, F-91297 Arpajon}
\altaffiltext{4}{LPGP, Universit\' e Paris Sud, F-91405 Orsay Cedex}
\altaffiltext{5}{LERMA, Observatoire de Paris, ENS, UPMC, UCP, CNRS, 5 Place Jules
Janssen, F-92195 Meudon}
\altaffiltext{6}{Theoretical Division, LANL, Los Alamos NM 87545,USA }
\altaffiltext{7}{AWE, Reading, Berkshire, RG7 4PR, UK}
\altaffiltext{8}{LULI, Ecole Polytechnique, CNRS, CEA, UPMC, F-91128 Palaiseau Cedex}

\begin{abstract}
The lifetime of solar-like stars, the envelope structure of more massive stars, and stellar acoustic frequencies largely depend on the radiative properties of the stellar plasma. Up to now, these complex quantities have been estimated only theoretically.  The development of the powerful tools of helio- and astero- seismology has made it possible to gain insights on the interiors of stars.  Consequently,  increased emphasis is now placed on knowledge of the monochromatic opacity coefficients.  Here we review how these radiative properties play a role, and where they are most important.  We then concentrate specifically on the envelopes of $\beta$ Cephei variable stars.   We discuss the dispersion of eight different theoretical estimates of the monochromatic opacity spectrum and the challenges we need to face to check these calculations experimentally. 
\end{abstract}


\section{Context}
After more than two decades of study, helioseismology has demonstrated the capability to probe the solar interior thanks to the seismic instruments of the SOHO satellite and ground-based networks
\citep{christensen04,guzik05,basu08,turck10a}.

The new space missions (SDO, PICARD) are oriented toward a better understanding of the Sun's internal magnetism and its impact on the Earth's atmosphere. In this context, the existence of a fossil field in the Sun's radiative zone is a missing bridge that motivate us to better estimate the competition between microscopic and macroscopic physics \citep{duez10,turck10b}. 

In addition to the international solar research community, more than 350 scientists from all over the world are working on the data from the COROT and Kepler spacecraft dedicated to exoplanets and astero-seismology (see http://smsc.cnes.fr/COROT for COROT and http://astro.phys.au.dk/KASC/ for Kepler). 
The development of these disciplines justifies ensuring the correctness of the many fundamental properties of the plasma (detailed abundances, reaction rates, opacity coefficients, etc.). 

The new High Energy Density laser facilities and the Z pinch machine now enable access to the radiative properties of plasmas of stellar type in the laboratory during pico and nano  second timescales; see \citep{bailey07,bailey09,turck09}, and references therein. This review is divided into three parts:
Section 2 is dedicated to solar-type stars, Section 3 to the more massive stars, and Section 4 to the study of the $\beta$ Cephei variable star case and a comparison between eight calculated iron opacity spectra. The challenges faced by the experiments are described in Section 5.

\section{Radiative properties of solar-type stars }
Stars produce both nuclear energy through nuclear reactions, and  thermal energy due to the internal temperature that exceeds everywhere some thousand degrees. The central plasma is heated to more than a dozen million K, and the heat is transported by convection or radiation phenomena. 

The low-mass stars ($<1.5 M_\odot$) have an internal radiative zone which disappears in the region where the photons interact so strongly with the partially ionized constituents that convection becomes the most efficient way to transport  the energy; see Figure 32 of \cite{turck93} or \cite{turck10c}.

In fact,  this boundary between radiative transport and convective transport differs from star to star depending on the star's mass, age, and composition. Hydrogen and helium are completely ionized in these radiative zones due to their  high temperature and density, but in the central conditions,  the heaviest elements (iron and above) are generally partially ionized and their contribution to the opacity coefficients is rather high, 20\% in the solar case; see equation 3 and figures in \cite{turck09}.  Nevertheless, their mass fraction is of the order of 10$^{-3}$. In the rest of the radiative zone, since both the temperature and density decrease with radius, by typically a factor 15 and 100 respectively, most of the heavy elements down to oxygen become partially ionized, so their relative contribution increases the Rosseland mean opacity $\kappa_R$ by a factor up to 100 at the base of the convection zone due to the multiple photon interactions on the electronic bound layers of more and more elements. 

As a direct consequence, the lifetime of a star of a given mass strongly depends on the amount of heavier elements (carbon to uranium) present in the star. The  total element mass fraction is generally divided into  X, Y, and Z, with X and Y corresponding to the mass fraction of hydrogen and helium, respectively, and Z corresponding to the remainder.  The lifetime of a 0.8 M$_\odot$ star with Z=0.001 is about 14 Gyr, while the lifetime of a 0.8 M$_\odot$ with Z=0.02 is about 22 Gyr.   When the heavy element contribution is small, the star contracts more quickly, the central temperature is higher, the opacity coefficients are different, and the nuclear reactions become more efficient.

This example shows the importance of verifying some detailed contributions to the total Rosseland mean opacity values to properly estimate how  $\kappa_R$ changes from star to star depending on its mass and on its detailed composition. 

The best known case is the Sun where where our efforts to incorporate new updates of the composition highlight the important effect of the heavy elements on the sound speed profile \citep{turck04, bahcall05,asplund09,turck10a,turck10b}. The difference between the standard model sound-speed predictions and the observed sound speed is larger than the error bars coming from the knowledge of the composition and from the known difference between OPAL and OP mean Rosseland opacity tables \citep{badnell05}. 

While for the solar conditions the Rosseland mean opacities agree in general among different calculations, the detailed comparison, element by element, shows rather large differences, as mentioned by Blancard et al. in Turck-Chi\`eze et al. 2010c.   So it is useful to compare opacity calculations of a specific element  to laboratory experiments; then the extrapolation to other stars of different composition will be more reliable. The first experiments have been performed at the Sandia Z machine to check the iron contribution at a temperature and density not far from those at the base of the solar convection zone  \citep{bailey07,bailey09}. Extensive measurements on the NIF or LMJ are needed for several well-chosen elements and temperatures between 193 and 400 eV to validate opacity calculations in the radiative zones of solar-like stars.  

Indeed the region just around the transition is clearly the most difficult region due to the interplay of microscopic and macroscopic phenomena (turbulence, rotation and magnetic fields).
In fact, in contrast with the rest of the radiative zone,  the boundary of the internal radiative zone depends directly on the detailed abundances of oxygen, iron, and neon and on the associated Rosseland mean opacity coefficients. It depends also indirectly on the turbulence of these layers associated with the fact that the stars are generally rotating. So the thermal instability produces a strong horizontal shear between a differentially latitudinal rotating convective zone and a reasonably flat rotation in the radiative zone
\citep{turck10b}. Moreover, the magnetic field in these layers probably plays a crucial role in the solar dynamo at the beginning of the 11-year cycle.

In the sub-surface layers of these stars,  radiation and convection are in competition and  the heavy elements are less important because most of the elements are neutral or even in molecules.  In the corona, the plasma is heated to a  million degrees, so the heavy elements play a crucial role,  but the very low density plasma is no longer in LTE (local thermodynamic equilibrium), so non-LTE calculations are needed to explain the observed lines \citep{colgan08}.  

\section{Radiative properties of more massive stars }
When the stellar mass increases, the central temperature increases and consequently carbon, oxygen, nitrogen can penetrate the Coulomb barrier.  In these conditions the hydrogen burning is no longer dominated by the weak interaction but by the strong interaction where these elements act as catalysts.  The nuclear burning becomes more rapid, and the lifetime of these stars decreases from billions to millions of years, the central temperature increases strongly and the radiative diffusion has no time to take place. The core of these stars  ($>1.5 M_\odot$) becomes convective,  while the envelope is radiative.  In the envelope, many elements are partially ionized, so the detailed calculation of the Rosseland mean opacity is even more complex because many electronic shells are open and the bound-bound processes are extremely important. 

Figure 1 illustrates the case of a 10 M$_\odot$ star. All the species including iron are totally ionized in the center of this star due to high temperature, 34 MK, and a relatively low density of 8 g/cm$^3$. The opacity is not as high (Fig 1b) as in lighter stars, but the luminosity due to the CNO cycle is so high that  $\nabla_{rad}$ is higher than $\nabla_{ad}$ (Fig 1a). When the ratio of gradients is greater than one, the region is convective; in the opposite case, the region is radiative. The rest of the star is radiative except in two small regions which correspond to the partially ionized contribution of helium  and hydrogen at low temperatures (see Fig 1d) and of the iron-group elements above 100,000 K (Fig 1c).  Figure 2 shows that the differences between OP and OPAL are particularly important for Ni, Mn and Cr and that the contribution of Cr and Ni could be very important in the case of OP.   Therefore, we have begun comparisons between different opacity calculations.

\begin{figure}[t]
\hspace{-0.8cm}\includegraphics[width=9cm]{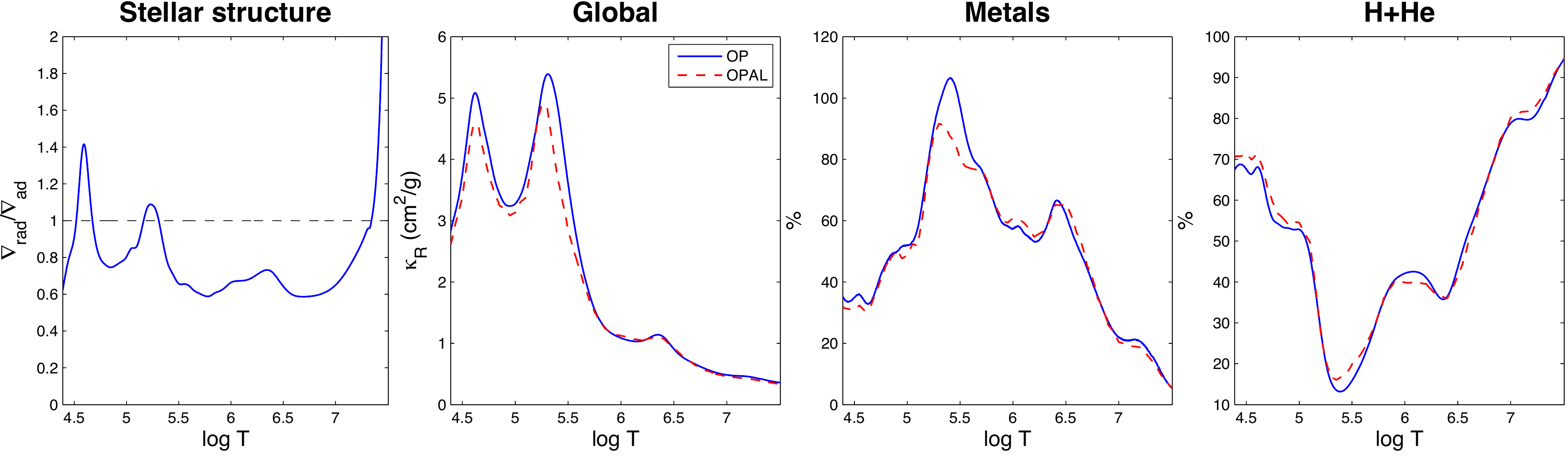}
\caption{%
a: Temperature gradient ratio between a radiative or convective region versus temperature for a 10 M$_\odot$ star. b: Temperature  dependence  of the total Rosseland mean opacity for OP calculation (solid blue line) and OPAL calculation (dashed red line). c: Relative contribution of the heavy elements to the total opacity. d: Relative contribution of the light elements to the total opacity.} \end{figure}

\begin{figure}[t]
\hspace{-0.4cm}\includegraphics[width=8.8cm]{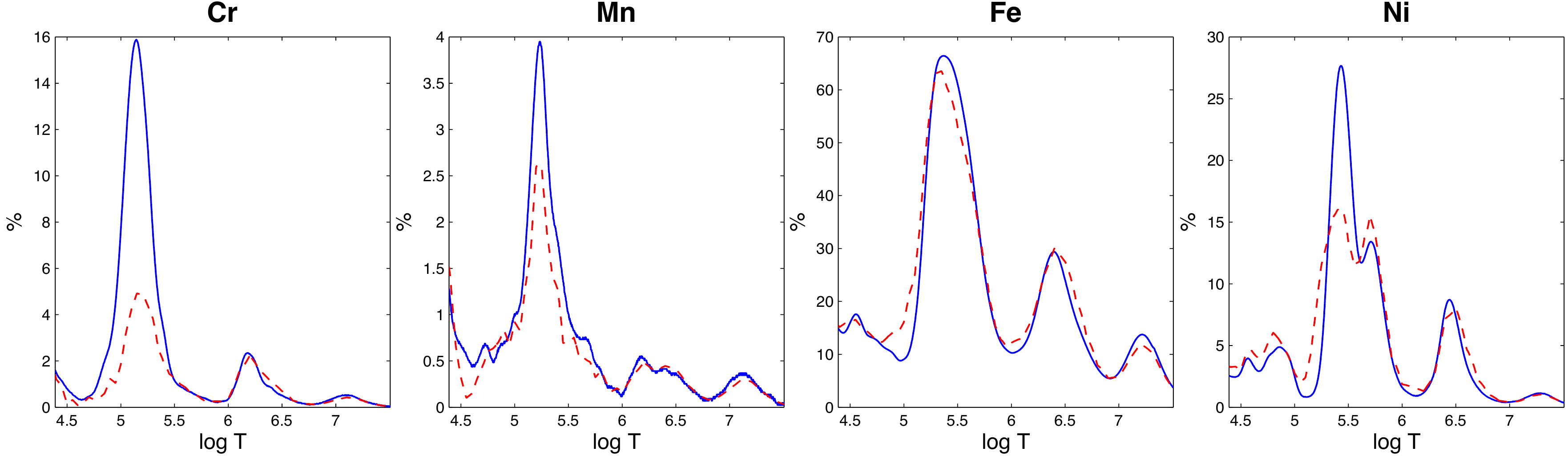}
\caption{%
Influence of the different elements of the iron group to the total opacity and comparison between OP and OPAL (same representation than Figure 1).} 
\end{figure}

Two other processes are considered in the stellar equations or in stellar pulsations. Both require a detailed description of  the sub-surface layers: the radiative acceleration and the $\kappa$ mechanism.

Radiative accelerations are  crucial quantities to study the diffusion of elements in stars and to understand the stellar photospheric abundances. They strongly depend on the atomic properties of the different ions, so they largely depend on the quality of the atomic data and on the knowledge of the ionization stage of the different elements. For a detailed description of this process, we refer to the book dedicated to G. Michaud, and in particular  to the paper of \cite{alecian05}. The expression for the radiative acceleration is also given in \cite{turck09}. 
This process must be taken into account in the description of stellar envelopes, in addition to the gravitational settling, turbulence and sometimes magnetic activity. For radiative accelerations, one needs to know the detailed monochromatic photon flux for all the processes (diffusion, bound-bound, bound-free...) and for each element. This information is not delivered by the OPAL consortium \citep{iglesias95} but is available for the OP calculations \citep{seaton04}. Some specific comparisons lead to 50\% differences in the radiative acceleration for the iron case in the opacity peak region mentioned in Figs. 1 and 2 \citep{delahaye05}.

It is important to resolve this discrepancy to understand the chemically peculiar stars and the thousands of pulsators observed with COROT and Kepler. The rapid variability of the above-mentioned opacities in the stellar envelope generates the so-called $\kappa$ mechanism and maintains the pulsation in classical Cepheids (due to partial ionization of helium) and for $\beta$ Cephei stars (due to the iron-group element ionization)  \citep{dupret03}.

\begin{figure}[t]
\includegraphics[width=7cm]{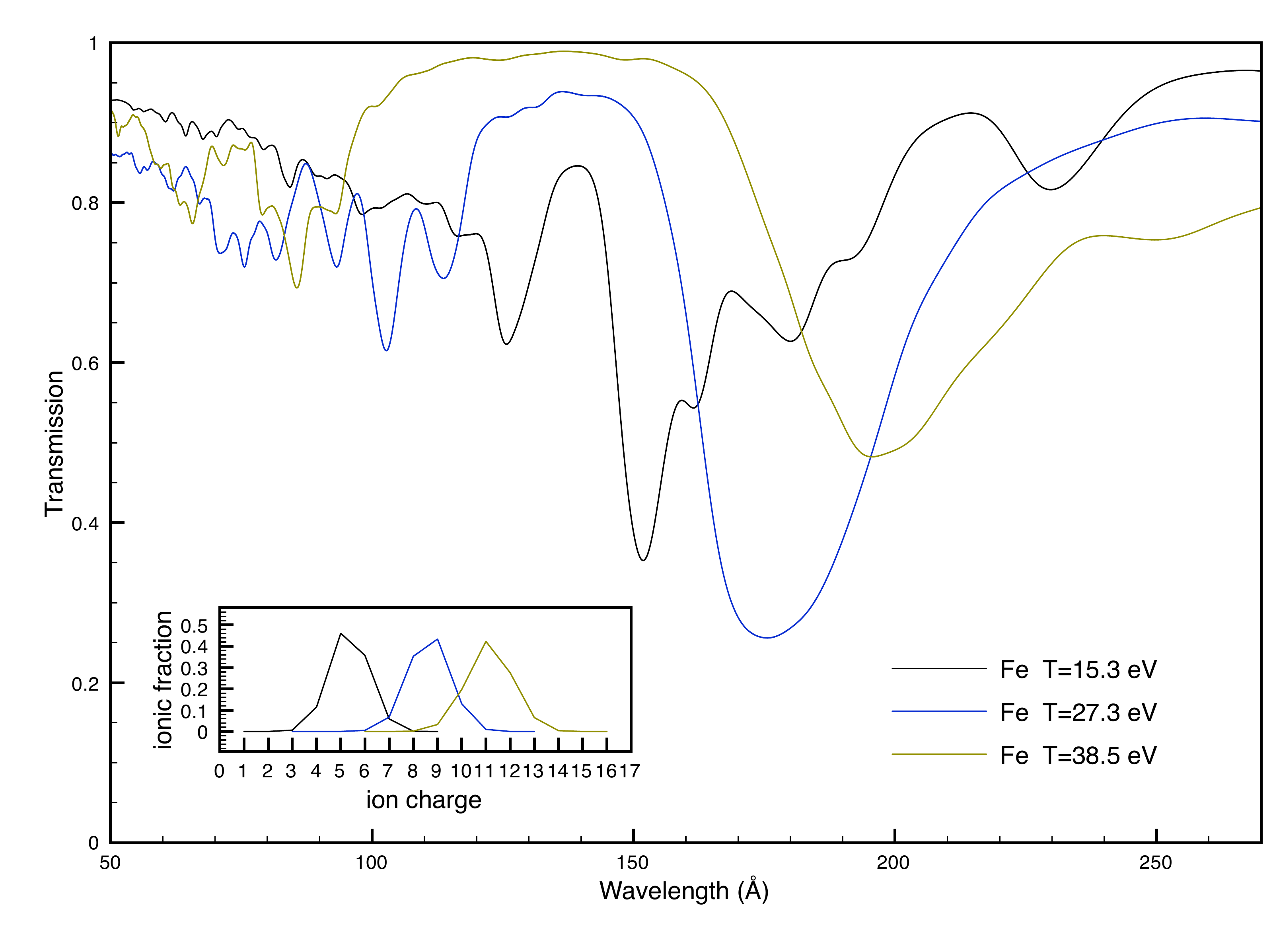}
\vspace{-0.5cm}
\caption{%
Iron transmission spectra corresponding to temperatures of 15.3, 27.3 and 38.3 eV and densities of 5.5 10$^{-3}$  g/cm$^3$, 3.4 10$^{-3}$  g/cm$^3$ and 2.6 10$^{-3}$  g/cm$^3$ respectively, estimated using the OP calculations. The whole wavelength range is useful for the Rosseland mean opacity determination ($\lambda$E = hc =12394 A.keV) and corresponds to the XUV spectrometer used for the experiment. The ionization charge distribution for each condition is superimposed.} 
\end{figure}

Calculations for asteroseismology must predict the theoretical acoustic modes in order to identify without ambiguity the observed modes and to understand the excitation of the modes. Several papers show the diffculties for $ \beta$ Cephei stars \citep{pamyatnykh99, briquet07, degroote10, daszynska10} and mention that an increase of opacity will improve the interpretation of the asteroseismic data.

\section{Theoretical opacity calculations} 
We investigate the plasma conditions found in the envelopes of  $\beta$ Cephei stars for improving the interpretation of the asteroseismic observations, without neglecting previous measurements of iron (and nickel) opacity spectra which have successfully led to a revision of opacities in the nineties, but not always in the range useful for astrophysics \citep{springer92,winhart95, chenais00,chenais01}. With this objective, we perform a comparison between different calculations of opacity spectra and an experiment to validate these calculations. 
 
The first step consists in defining a condition compatible with a laboratory experiment. The iron-group peak in the envelope of $\beta$ Cephei stars corresponds to temperatures of 200,000 K (17.2 eV, 1 eV= 11,600K) and densities around 4 10$^{-7}$  g/cm$^3$. This density is too low to be investigated in a laboratory experiment. So we have first looked for conditions which maintain the same distribution of ionization stage for iron.  

\begin{figure}[t]
\vspace{-3.5cm}
\hspace{-1.3cm}\includegraphics[width=9.7cm]{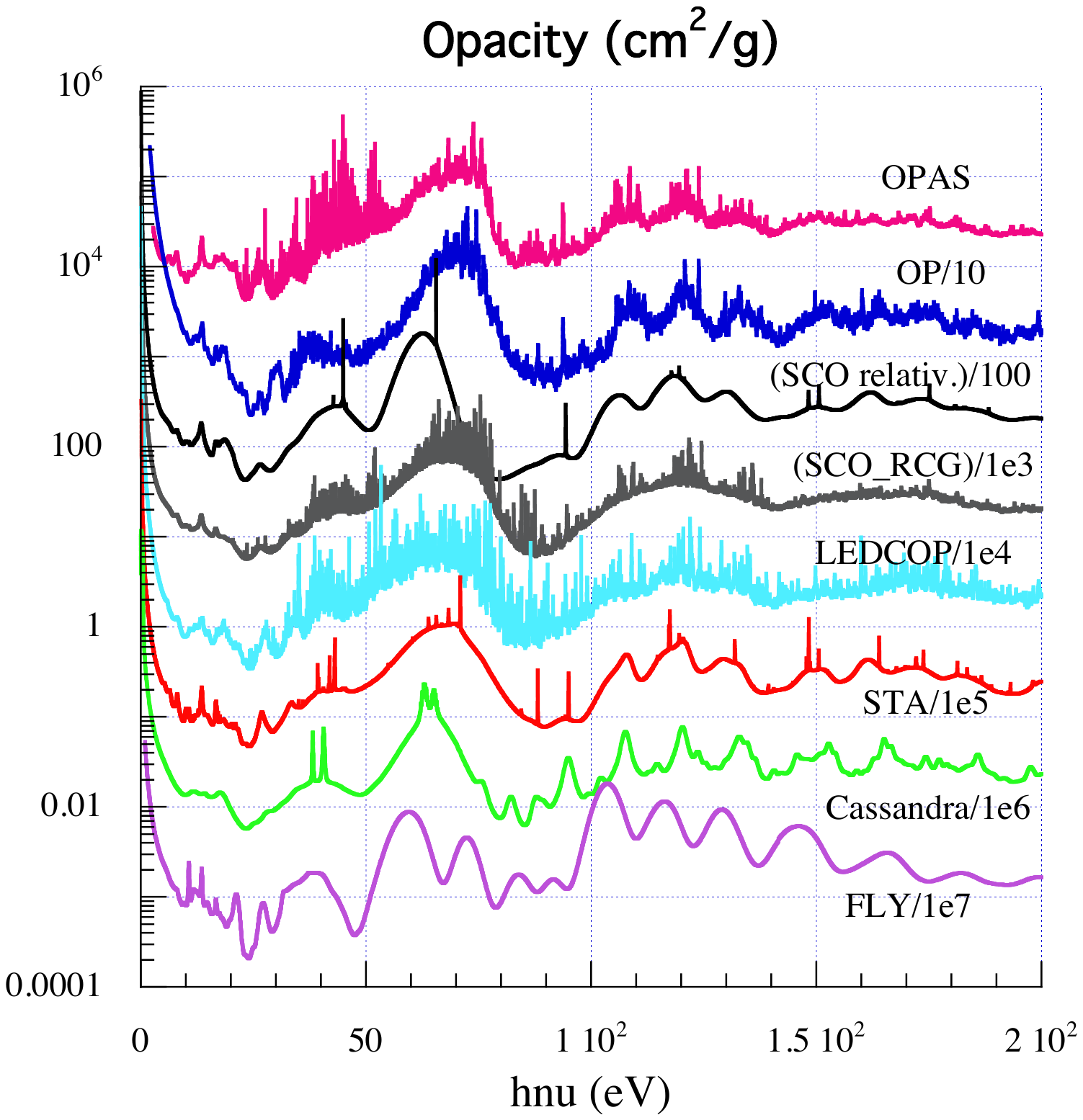}
\vspace{-3.5cm}
\caption{%
Comparison of eight theoretical monochromatic spectra for the mean conditions T= 27.3 eV and $\rho$ = 3.4 10$^{-3} g/cm^3$ shown in Figure 3. The hypotheses of these computations are explained briefly in the text.} 
\end{figure}

Figure 3 shows the spectral transmission of photons  $T_\nu$ through a heated iron foil of thickness $\epsilon$,  $T_\nu= exp^{-\kappa_\nu \rho \epsilon}$, where $\kappa_\nu$ is the opacity per mass unit in cm$^2$/g.  Transmission is the quantity  measured in the experiment. The mean spectrum is obtained using the OP tables, for the same distribution of the iron ionization charge (about 9 for T= 320,000 K or 27.3 eV, $\rho$ = 3.4 10$^{-3}$  g/cm$^3$). Two other spectra also are shown, for a variation of temperature of about 40\%. In these cases,  the distribution of charge is displaced by two or three units. The figure shows that the spectra vary strongly with temperature. In the region where the $\kappa$ mechanism is important, the Rosseland mean opacity varies rapidly for this reason, so measurement at different temperatures will help to check the whole calculation of the pulsation mode excitation.

Figure 4 shows the opacity spectra versus the photon energy obtained by the different calculations generally used to describe these plasmas. Clear differences can be seen in different parts of the spectral range. The corresponding Rosseland mean values show differences of up to 40\%, in particular for the OP value, but most of them agree within 15\% despite the very different calculational approaches. We notice also some variation of the mean ionization from one calculation to another of about 0.5 units, which is not so small if one considers the rapid variation of the opacity with the degree of ionization, but which cannot explain all of the differences. A detailed analysis of these calculations will appear in a more detailed paper. We give here a short description of the assumptions. 

Based on the average atom model SCAALP \citep{blancard04}, the OPAS code combines detailed configuration and level accounting treatments to calculate radiative opacity of plasmas in LTE. The bound-bound opacity can be calculated using detailed line accounting and/or statistical methods. The bound-free
opacity is evaluated neglecting the configuration level splitting. To improve the
accuracy of opacities into the complex regime of warm dense matter, where plasma
and many-body effects can be important, the free-free opacity is obtained by
interpolating between the Drude-like opacity and the opacity derived from the Kramers
formula including a Gaunt factor and an electron degeneracy effect correction. Photon scattering by free electrons includes collective effects as well as relativistic corrections.
 
OP (Opacity Project) is an on-line atomic database used in astrophysics \citep{seaton04,seaton07}.
TOPbase contains the most complete dataset of LS-coupling term energies, f-values and photoionization cross sections for the most abundant ions (Z=1-26) in the universe. They have been computed in the close-coupling approximation by means of the R-matrix method with innovative asymptotic techniques. TOPbase also contains large datasets of f-values for iron ions with configurations $\rm3s^x3p^x3d^x$, referred to as the PLUS-data, computed with the atomic structure code SUPERSTRUCTURE.

The atomic physics SCO code \citep{blenski97,blenski00} is based on the superconfiguration approach proposed by \cite{barshalom89}  to calculate high-Z element photoabsorption in LTE conditions. Here calculations have been performed including non-relativistic and relativistic contributions.

SCO-RCG \citep{porcherot10} is a hybrid opacity code which combines the statistical super-transition-array approach and fine-structure calculations. Criteria are used to select transition arrays which are removed from the super-configuration statistics, and replaced by a detailed line-by-line treatment. The data required for the calculation of the detailed transition arrays (Slater, spin-orbit and dipolar integrals) are obtained from the super-configuration code SCO (see above), providing in this way a consistent description of the plasma screening effects on the wave functions. Then, the level energies and the lines (position and strength) are calculated by the routine RCG of R.D. Cowan's atomic structure code \citep{cowan81}.

\indent
The Los Alamos Light Element Detailed Configuration OPacity code, LEDCOP \citep{magee95},
uses a basis set of LS terms based on semi-relativistic Hartree-Fock 
atomic structure calculations (plus a mixture of LS terms and unresolved transition arrays, some of 
which are treated by random lines for the most complex ion stages) to calculate opacities for elements with Z $<$ 31.
Each ion stage is considered in detail and interactions with the plasma
are treated as perturbations.  Calculations are done in local thermodynamic
equilibrium (LTE) and only radiative processes are included.

In  the STA opacity code \citep{barshalom89} the total transition array of a specific single-electron transition, including all possible contributing configurations, is described by only a small number of super-transition-arrays (STAs). Exact analytic expressions are given for the first few moments of an STA. The method is shown to interpolate smoothly between the average-atom (AA) results and the detailed configuration accounting that underlies the unresolved transition array (UTA) method. Each STA is calculated in its own optimized potential, and the model achieves rapid convergence in the number of STAs included.

CASSANDRA is an AWE (Atomic Weapons Establishment, Aldermaston, UK) opacity code \citep{crowley01} used to model plasma in local thermal equilibrium. CASSANDRA's self-consistent field calculation uses the local density approximation for bound states and has a free electron contribution based upon the Thomas-Fermi model. The Cassandra data are ©British Crown Owned Copyright 2010/MOD.

FLYCHK is a free access atomic spectroscopy code  \citep{chung05} used to generate atomic level populations and charge state distributions and synthetic spectra for low-Z to mid-Z elements under LTE or NLTE conditions http://nlte.nist.gov/FLY/.
\section{Experimental challenges}
\begin{figure}[t]
\hspace{-0.2cm}\includegraphics[width=8.cm]{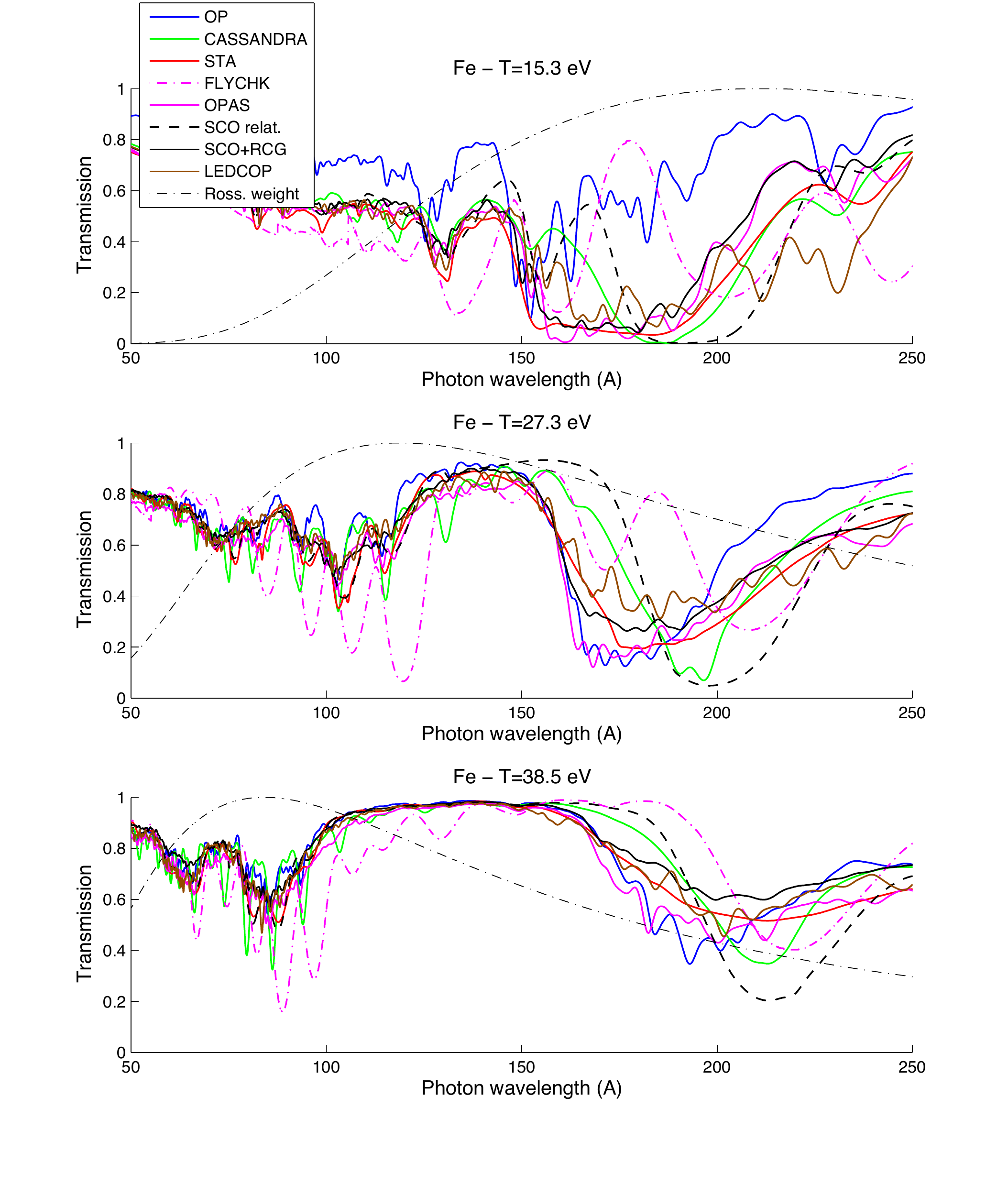}
\caption{Transmission spectra of the calculations for conditions mentioned in Figure 3 and a XUV spectrometer resolution of 1 eV.  Superimposed is the weighted shape of the Rosseland mean opacity (dot-dash line). }
\end{figure}

The new spectral opacity measurements have been performed on LULI 2000 www.luli.polytechnique.fr/ with two lasers in 2010. 
 A nanosecond laser delivers an energy between 30-500 J in a 500 ps duration pulse. This laser is used to irradiate a gold cavity (hohlraum) on which an absorbing foil of the considered element is deposited and heated. After a required delay to get the proper density and temperature, a picosecond laser interacts with a backlighter to produce x rays which interact with the formed plasma: for a description of the experimental setup, see \cite{loisel09}. We detect the transmitted opacity spectrum on a streak camera placed behind a newly designed XUV-ray spectrometer. Obtaining high-quality  measurements required addressing many challenges based on experience with previous measurements.  A detailed analysis of the previous experiments and their difficulties is found in \cite{chenais02,bailey09}. We summarize some of these relevant for astrophysical applications:

- One needs to form a plasma in LTE. This requirement supposes a good simulation of the whole experiment to probe the foil in the best conditions. The rapid expansion of this foil during heating is limited by placing the foil between two thin samples of a low-Z material (typically carbon).

- We measure the foil temperature with a $\mu$Dmix and limit the temperature gradient by placing the foil between two cavities. The temperatures depend on the geometry of the experiment, the position of the foil, and the correctness of the simulation.

- The resolution of the XUV spectrometer is important to discriminate between calculations; it  varies with the wavelength  and one must avoid saturation. 

Figure 5 shows the transmission obtained for the different calculations including a spectrometer resolution of 1 eV for the the conditions mentioned in Figure 3;  the differences must be distinguishable by the experiment. Notice the distinct behavior of the OP calculation at low energy and some saturation of the spectra which illustrates the difficulties of exploring the low-energy regime.

To reduce the difficulties noted above, we measure the spectra of 3 or 4 neighboring elements (chromium, iron, nickel, copper) using two thicknesses for each of them.  Moreover we use two cavities to heat the foil on the two faces to limit the temperature gradient. 

\vspace{.3cm}
In summary, this study shows that it would be of interest to develop complete opacity tables including photon energy spectra with the best physics available for asteroseismology.  It is important to improve the coverage of some crucial regions for interpreting the space observations.  Some special validations by laboratory experiments are required to validate "the best physics".
\makeatletter
\let\clear@thebibliography@page=\relax
\makeatother


\begin{thebibliography}{}
\bibitem[{{Alecian}(2005)}]{alecian05} Alecian, G. 2005, EAS Pub. Series, 17, 33
\bibitem[{{Asplund} {et al.}(2009)}]{asplund09} Asplund, M., Grevesse, N., and Scott, P. 2009, ARA\&A, 47, 481
\bibitem[Badnell et al.(2005)]{badnell05} Badnell, N. R. et al. 2005, \mnras, 360, 458.  
\bibitem[{{Bahcall} {et al.}(2005)}]{bahcall05}Bahcall, J.~N., Serenelli, A. M. \& Basu, S. \ 2005, \apjl, 621, L85
\bibitem[{{Bailey} {et al.} (2007)}]{bailey07}Bailey, J. et al., 2007, Phys. Rev. Lett. 99, 5002
\bibitem[{{Bailey} {et al.}(2009)}]{bailey09}Bailey, J. et al., 2009, Phys. of Plasmas, 16, 058101
\bibitem[{{Bar-Shalom} {et al.} (1989)}]{barshalom89}Bar-Shalom, A., Oreg, J., Goldstein, W. H., Schwarts, D., Zigler, Z. 1989, Phys. Rev. A, 40,  3183 
\bibitem[Basu and Antia (2008)]{basu08} Basu, S., Antia, H. M. 2008, Phys. Rep. 457, 217
\bibitem[{{Blancard \& Faussurier} (2004)}]{blancard04} Blancard, C. and Faussurier, G. 2004
PRE 69, 016409
\bibitem[{{Blenski,Grimaldi \& Perrot}  (1997)}]{blenski97}Blenski, T., Grimaldi, A. and Perrot, P. 1997, Phys. Rev. E 55, R4889 
\bibitem[{{Blenski} {et al.} (2000)}]{blenski00} Blenski, T., Grimaldi, A.,  Perrot, F.,  J. 2000, Quant. Spectrosc. Radiat. Transf., 65, 91
\bibitem[{{Briquet} {et al.} (2007)}]{briquet07} Briquet, M. et al. 2007, MNRAS, 381, 1482
\bibitem[{{Chenais-Popovics} {et al.} (2000)}]{chenais00}  Chenais-Popovics, C. et al. 2000 ApJ Suplem. Series. 127,  175
\bibitem[{{Chenais-Popovics} {et al.} (2001)}]{chenais01} Chenais-Popovics, C., Fajardo, M., Thais, F. et al. 2001, J. Quant. Spectrosc. Radiat.
Transf.,  71, 249.
\bibitem[{{Chenais-Popovics} (2002)}]{chenais02}Chenais-Popovics, C. 2002, Laser and Particle Beams 20, 291
\bibitem[Christensen-Dalsgaard (2004)]{christensen04} Christensen-Dalsgaard, J. 2004, Sol. Phys., 220, 137
\bibitem[Chung et al.(2005)]{chung05}Chung, H.-K., Chen, M. H., Morgan, W. L., Ralchenko, Y., Lee, R. W. 2005, HEDP 1, 3
\bibitem[Colgan et al.(2008)]{colgan08} Colgan, J., Abdallah, J., Jr., Sherrill, M. E., Foster, M., Fontes, C.J. and Feldman, U. 2008, \apj,
    689, 585
\bibitem[Cowan (1981)]{cowan81} Cowan, R.D. 1981, The Theory of Atomic Structure and Spectra, University of California Press 
\bibitem[Crowley et al. (2001)]{crowley01} Crowley, B.J.B.  et al.  2001, JQRST, 71, 257
\bibitem[Daszynska-Daszkiewicz \& Walczak (2010)]{daszynska10} Daszynska-Daszkiewicz, J. \& Walczak, P. 2010, \mnras, 403, 496   
\bibitem[Delahaye \& Pinsoneault (2006)]{delahaye05} Delahaye, F. \& Pinsonneault, M. 2006 ApJ,  625, 563
\bibitem[{{Degroote} {et al.}(2010)}]{degroote10} Degroote, P. et al. 2009  A\&A,   506, 111 
\bibitem[Duez, Mathis \& Turck-Chi\`eze (2010)]{duez10} Duez, V., Mathis, S. and Turck-Chi\`eze, S.  2010, \mnras, 402, 271
\bibitem[Dupret (2003)]{dupret03} Dupret, M. A. 2003, thesis, University of Liege, Belgium
\bibitem[Iglesias \& Rogers (1995)]{iglesias95} Iglesias, C.A.  \& Rogers, F. J. 1995,  ApJ,  443, 469
\bibitem[Guzik, Watson \& Cox (2005)]{guzik05} Guzik, J.A., Watson, L.S. \& Cox, A N. 2005, ApJ, 627, 1049 
\bibitem[{{Loisel} {et al.}(2009)}]{loisel09} Loisel, G. et al. \ 2009, HEDP, 5, 173
\bibitem[{{Magee} {et al.}(1995)}]{magee95} Magee, N.H., et al. 1995, in ASP Conf. Ser. 78, 
51
 \bibitem[{{Pamyantnykh}(1999)}]{pamyatnykh99} Pamyatnykh, A.A. 1999, Acta Astronomica, 49, 119 
 \bibitem[{{Porcherot et al.} (2010)}]{porcherot10}  Porcherot, Q., Gilleron, F.,  Pain, J.C., and  Blenski, T. 2010, to be published. 
\bibitem[Seaton \& Badnell (2004)]{seaton04}  Seaton, M.J. \& Badnell, N.R. 2004, MNRAS, 354,  457
\bibitem[Seaton (2007)]{seaton07}  Seaton M.J. 2007, MNRAS, 382, 245
\bibitem[{{Springer} { et al.}  (1992)}]{springer92} Springer, P.T. et al. 1992, \prl, 69, 3735 
\bibitem[{{Turck-Chi{\`e}ze} {et al.}(1993)}]{turck93}Turck-Chi{\`e}ze, S., et al. 1993, Phys. Rep., 230, 59-235
\bibitem[{{Turck-Chi{\`e}ze} {et al.}(2004)}]{turck04} Turck-Chi{\`e}ze, S., et al. \ 2004, \prl, 93, 211102  
\bibitem[{{Turck-Chi{\`e}ze} {et al.}(2009)}]{turck09}Turck-Chi{\`e}ze, S., et al. 2009,  HEDP, 5, 132
\bibitem[{Turck-Chi\`eze \& Couvidat (2010a)}]{turck10a}  Turck-Chi\`eze, S. and Couvidat, S. 2010, Rep. Prog. Phys., submitted
\bibitem[{{Turck-Chi\`eze} {et al.}(2010b)}]{turck10b}  Turck-Chi\`eze, S., Palacios, A., Marques, J., Nghiem, P.A.P. 2010, \apj, 715, 1539
\bibitem[{{Turck-Chi\`eze} {et al.}(2010b)}]{turck10c}  Turck-Chi\`eze, S.,  Loisel, G., Gilles, D. et al. 2010c, SOHO24, J. Phys.: Conf. Ser., in press
\bibitem[{{Winhart} {et al.} (1995)}]{winhart95}Winhart, G. et al. 1995, JQSRT, 54, 37
 \end{thebibliography}
 \end{document}